\documentclass[pdflatex,sn-nature]{sn-jnl}% Style for submissions to Nature Portfolio journals
\usepackage{graphicx}%
\usepackage{multirow}%
\usepackage{amsmath,amssymb,amsfonts}%
\usepackage{amsthm}%
\usepackage{mathrsfs}%
\usepackage[title]{appendix}%
\usepackage{xcolor}%
\usepackage{textcomp}%
\usepackage{manyfoot}%
\usepackage{booktabs}%
\usepackage{algorithm}%
\usepackage{algorithmicx}%
\usepackage{algpseudocode}%
\usepackage{listings}%
\usepackage{comment}
\theoremstyle{thmstyleone}%
\theoremstyle{thmstyletwo}%

\theoremstyle{thmstylethree}%

\begin{document}

\title[SSL Benchmark]{A Clinical Benchmark of Public Self-Supervised Pathology Foundation Models}

%%=============================================================%%
%% GivenName	-> \fnm{Joergen W.}
%% Particle	-> \spfx{van der} -> surname prefix
%% FamilyName	-> \sur{Ploeg}
%% Suffix	-> \sfx{IV}
%% \author*[1,2]{\fnm{Joergen W.} \spfx{van der} \sur{Ploeg} 
%%  \sfx{IV}}\email{iauthor@gmail.com}
%%=============================================================%%

\author*[1,2]{\fnm{Gabriele} \sur{Campanella}}\email{gabriele.campanella@mssm.edu}
\author[1,2]{\fnm{Shengjia} \sur{Chen}}
\author[1,2]{\fnm{Ruchika} \sur{Verma}}
\author[3]{\fnm{Jennifer} \sur{Zeng}}
%\equalcont{These authors contributed equally to this work.}
\author[3]{\fnm{Aryeh} \sur{Stock}}
\author[3]{\fnm{Matt} \sur{Croken}}
\author[3]{\fnm{Brandon} \sur{Veremis}}
\author[4]{\fnm{Abdulkadir} \sur{Elmas}}
\author[4]{\fnm{Kuan-lin} \sur{Huang}}
\author[3]{\fnm{Ricky} \sur{Kwan}}
\author[3]{\fnm{Jane} \sur{Houldsworth}}
\author[5]{\fnm{Adam J.} \sur{Schoenfeld}}
\author[6]{\fnm{Chad} \sur{Vanderbilt}}
%\author*[1,2]{\fnm{Thomas J.} \sur{Fuchs}}\email{thomas.fuchs.ai@mssm.edu}

\affil*[1]{\orgdiv{Windreich Department of AI and Human Health}, \orgname{Icahn School of Medicine at Mount Sinai}, \orgaddress{\city{New York}, \postcode{10029}, \state{NY}, \country{United States}}}
\affil[2]{\orgdiv{Hasso Plattner Institute at Mount Sinai}, \orgname{Icahn School of Medicine at Mount Sinai}, \orgaddress{\city{New York}, \postcode{10029}, \state{NY}, \country{United States}}}
\affil[3]{\orgdiv{Department of Pathology}, \orgname{Icahn School of Medicine at Mount Sinai}, \orgaddress{\city{New York}, \postcode{10029}, \state{NY}, \country{United States}}}
\affil[4]{\orgdiv{Department of Genetics and Genomics}, \orgname{Icahn School of Medicine at Mount Sinai}, \orgaddress{\city{New York}, \postcode{10029}, \state{NY}, \country{United States}}}
\affil[5]{\orgdiv{Department of Medicine}, \orgname{Memorial Sloan Kettering Cancer Center}, \orgaddress{\city{New York}, \postcode{10065}, \state{NY}, \country{United States}}}
\affil[6]{\orgdiv{Department of Pathology}, \orgname{Memorial Sloan Kettering Cancer Center}, \orgaddress{\city{New York}, \postcode{10065}, \state{NY}, \country{United States}}}

%%==================================%%
%% Sample for unstructured abstract %%
%%==================================%%

\abstract{The use of self-supervised learning (SSL) to train pathology foundation models has increased substantially in the past few years. Notably, several models trained on large quantities of clinical data have been made publicly available in recent months. This will significantly enhance scientific research in computational pathology and help bridge the gap between research and clinical deployment. With the increase in availability of public foundation models of different sizes, trained using different algorithms on different datasets, it becomes important to establish a benchmark to compare the performance of such models on a variety of clinically relevant tasks spanning multiple organs and diseases.
In this work, we present a collection of pathology datasets comprising clinical slides associated with clinically relevant endpoints including cancer diagnoses and a variety of biomarkers generated during standard hospital operation from two medical centers. We leverage these datasets to systematically assess the performance of public pathology foundation models and provide insights into best practices for training new foundation models and selecting appropriate pretrained models.
}


\maketitle

\section{Introduction}\label{sec:intro}
Artificial Intelligence (AI) is revolutionizing the medical field. The introduction of deep learning~\cite{lecun_deep_2015} has greatly accelerated the development of predictive models for high-dimensional data modalities such as images and text that are not readily amenable to classical machine learning algorithms. Convolutional neural networks (CNNs) and vision transformers~\cite{dosovitskiy_image_2020} (ViTs) have been used to solve numerous problems using supervised learning and have enabled the training of predictive models for a variety of tasks with high performance. Recently, the development of self-supervised learning (SSL) algorithms has marked a paradigm shift by enabling the training of deep neural networks on very large unlabeled datasets, yielding results on par with supervised learning strategies. Large neural networks trained this way can be described as foundation models that can be used for a wide variety of downstream tasks with little to no fine-tuning. Despite the great successes in the computer vision and natural language fields, SSL algorithms and foundation models are still in their infancy in the medical domain. One of the main reasons is the lack of medical datasets and the necessary computing infrastructure which makes large-scale SSL experiments only possible at large well-funded institutions.

In pathology, the lack of data is even more acute due to the still low adoption of digital pathology. Additionally, digital whole slide images (WSI) are orders of magnitude larger than other image modalities, with resolutions of tens to hundreds of thousands of pixels in each dimension. This poses challenges in terms of the methods used to analyze the images and the hardware requirements to effectively perform experiments. A common strategy to analyze these images is to divide the slide into small tiles or patches and encode them using a deep neural network, expressing the slide as a list of feature vectors and thus reducing the dimensionality of the slide by multiple orders of magnitude. In a second step, the feature vectors are aggregated using a neural network to obtain a slide-level representation. The first step is by far the most computationally expensive, while the second step requires much fewer resources. This is why most studies in computational pathology rely on already existing pretrained encoders, usually trained on natural images and not WSIs. There is a need for strategies that enable training of encoders directly on pathology images, and SSL lends itself well for this task as it does not require any sort of labels and does allow for the training of a pathology foundation model on large datasets. SSL for pathology has recently received lots of attention, and there are many academic and non-academic efforts to build a general-purpose pathology foundation model (Table~\ref{tab:works}).

Wang et al.~\cite{wang_transformer-based_2022} proposed SRCL, an SSL method based on MoCo v3~\cite{chen_empirical_2021}, along with CTransPath, a model architecture that combines convolutional layers with the Swin Transformer~\cite{liu_swin_2021} model. They trained their model on 15.6 million tiles from 32,220 slides from the TCGA~\cite{the_cancer_genome_atlas_research_network_cancer_2013} and PAIP datasets spanning 25 anatomic sites and over 32 cancer subtypes. The downstream performance was assessed on patch retrieval, supervised patch classification, weakly-supervised WSI classification, mitosis detection, and colorectal adenocarcinoma gland segmentation. Methodological advances include the introduction of a strategy to sample positive examples for the contrastive learning approach, and the hybrid convolutional-transformer model architecture.

Filiot et al.~\cite{filiot_scaling_2023} analyzed the performance of iBOT~\cite{zhou_ibot_2021}, an SSL framework that combines masked image modeling and contrastive learning, on histology data. They trained several ViT models on a dataset consisting of up to 43.3 million tiles from 6,093 TCGA slides of 13 anatomic sites. They assessed the performance of learned features on 17 downstream tasks across seven cancer indications including tile-level and slide-level tasks for subtype, genomic alteration, and overall survival prediction.

Chen et al.~\cite{chen_general-purpose_2023,chen_towards_2024} introduced UNI, a ViT-large model trained on 100,000 proprietary slides using the DINOv2~\cite{oquab_dinov2_2023} SSL algorithm. The pretraining dataset they used included 100 million tiles from 20 major tissue types. They evaluated the downstream performance across 33 tasks, which included tile-level tasks such as classification, segmentation, retrieval, as well as slide-level classification tasks.

Vorontsov et al.~\cite{vorontsov_virchow_2023} introduced Virchow, a ViT-huge model trained on 2 billion tiles from almost 1.5 million proprietary slides with DINOv2~\cite{oquab_dinov2_2023}. Slides were included from 17 tissue types and the performance on downstream tasks was evaluated using tile-level and slide-level benchmarks, encompassing tissue classification and biomarker prediction.

Campanella et al.~\cite{campanella_computational_2023} compared the performance of masked autoencoders~\cite{he_masked_2021} (MAE) and DINO~\cite{caron_emerging_2021} using over 3 billion tiles sourced from more than 423,000 pathology slides. The models were evaluated on six clinical tasks spanning three anatomical sites and two institutions. Their results showed the superiority of the DINO algorithm for pathology foundation model pretraining.

Dippel et al.~\cite{dippel_rudolfv_2024} introduced RudolfV, a model that integrates pathologist expertise, semi-automated data curation, and a diverse dataset from over 15 laboratories. Their dataset comprised 134,000 slides from 34,000 cases, representing a broad spectrum of histological samples with various fixation, staining, and scanning protocols from laboratories across the EU and US. Additionally, semantically similar slides and tissue patches were grouped to optimize data sampling for training, and stain-specific data augmentation was applied. %RudolfV demonstrates superior performance compared to existing state-of-the-art models, excelling in benchmarks focused on tumor microenvironment profiling, biomarker evaluation, and reference case search, while exhibiting favorable robustness properties.

Xu et al.~\cite{xu_whole-slide_2024} introduced Prov-GigaPath, that was created by tile-level pretraining using DINOv2~\cite{oquab_dinov2_2023}, followed by slide-level pretraining using a masked autoencoder~\cite{he_masked_2021} and LongNet~\cite{ding2023longnet}. This model was pretrained on 1.3 billion tiles derived from 171,189 WSIs comprising H\&E-stained and immunohistochemistry (IHC) slides from Providence Health and Services. These WSIs originated from over 30,000 patients encompassing 31 tissue types. Prov-GigaPath was evaluated on 17 genomic prediction tasks and 9 cancer subtyping tasks using both Providence and TCGA~\cite{cancer2014comprehensive} data. %Prov-GigaPath achieved state-of-the-art performance on 25 out of 26 tasks and the model's superior performance across various digital pathology tasks underscores the importance of real-world data and whole-slide modeling.

\begin{table}[h]
\caption{A summary of recently published pathology foundation models. 
MGB: Mass General Brigham,
%MGH: Massachusetts General Hospital, BWH: Brigham and Women's Hospital, 
MSKCC: Memorial Sloan Kettering Cancer Center, 
MSHS: Mount Sinai Health System,
PHS:  Providence Health and Services}\label{tab:works}
\begin{tabular*}{\textwidth}{@{\extracolsep\fill}lrllrrr}
\toprule
& Param. & Algorithm & Training Data & Tiles & Slides & Organs \\
Model & (M) & & Source & (M) & (K) &   \\
\midrule
CTransPath~\cite{wang_transformer-based_2022} & 28 & SRCL & TCGA, PAIP & 16 & 32 & 25 \\
Phikon~\cite{filiot_scaling_2023} & 86 & iBOT & TCGA & 43 & 6 & 13 \\
UNI~\cite{chen_towards_2024} & 303 & DINOv2 & MGB & 100 & 100 & 20 \\
Virchow~\cite{vorontsov_virchow_2023} & 631 & DINOv2 & MSKCC & 2,000 & 1,488 & 17 \\
Campanella et al.~\cite{campanella_computational_2023} & 22 & DINO & MSHS & 1,600 & 423 & 42 \\
Campanella et al.~\cite{campanella_computational_2023} & 303 & MAE & MSHS & 3,200 & 423 & 42 \\
Rudolf-V~\cite{dippel_rudolfv_2024} & 304 & DINOv2 & Multicenter & 1,200& 134 & 14 \\
Prov-GigaPath~\cite{xu_whole-slide_2024} & 1,135 & DINOv2 & PHS & 1,300 & 171 & 31\\
\botrule
\end{tabular*}
\end{table}

It is becoming abundantly clear that using SSL to train image encoders on unlabeled pathology data is superior to relying on models pretrained on other domains such as natural images~\cite{campanella_computational_2023}. While SSL trained pathology models hold immense potential, there are still some challenges that need to overcome before pathology foundation models can be used reliably in clinical workflows.
One consideration is that datasets used to train pathology models are still relatively small compared to other domains, in particular natural images, especially when considering the number of slides or cases. Since each pathology slide can contain tens of thousands of tiles, it is possible to generate large number of tiles from a small number of slides. Thus, it is essential to consider not only the number of tiles or slides used, but also other metrics of tissue heterogeneity such as anatomic sites and organ inclusion. Given the evidence from the natural language and vision domains that larger datasets and higher capacity models will produce better performance especially in the SSL setting, training on larger pathology datasets should be a priority. Recent works show progress in this respect as the digitization of pathology data becomes more prevalent.
Most importantly, the downstream performance of SSL models for pathology should be assessed on clinically derived data, preferably from multiple institutions, for clinically relevant tasks such as diagnostic assessment, biomarker prediction, and outcome prediction. Tile-based predictions, organ classification, coarse segmentation, captioning, retrieval, and visual question answering (VQA) are valuable scientific explorations, but less relevant in the clinical setting. This effect is compounded by the use of curated public datasets which may not be suited for assessing generalization to real world data. It should be noted that progress in this regard is being made and a trend towards the use of more clinical data in recent publications can be observed. Yet, there is still a lack of a systematic comparison of current models on a wide variety of clinical tasks.

In the present work we overcome this limitation by introducing a clinical benchmark dataset which is used to systematically compare public pathology foundation models. The dataset consists of clinical data generated during standard hospital operations from two different institutions. It includes three broad task types (disease detection, biomarker prediction, and treatment outcome prediction), and a wide range of disease indications and anatomic sites. This is an ongoing effort. As new foundation models are published and additional datasets are added to our benchmarks, we will regularly update our findings to provide the community with a comprehensive view of the state of foundation models in computational pathology. The live benchmark can be found on \href{https://github.com/fuchs-lab-public/OPAL/tree/main/SSL_benchmarks}{GitHub}.

\section{Method}\label{sec:method}
In the SSL literature, the performance of downstream tasks is frequently assessed by training a linear classifier (linear probing) on top of features extracted by a frozen encoder, or via zero-shot approaches such as k-NN. For pathology slides, there is no direct way to translate these approaches without having tile-level annotations. Instead, it is common practice to train a slide-level aggregator. For this purpose we chose the popular Gated MIL Attention (GMA) model~\cite{ilse_attention-based_2018} with a linear classifier on top. Since GMA does not consider the spatial distribution of tiles over the slide in its prediction, it is a simple method to test the expressiveness of the feature space generated by the SSL pretraining.

For each slide, tissue tiles were extracted at 20x magnification (0.5 microns per pixel, MPP) and embedded into a feature representation using a specific foundation model. Each slide is then converted to a 2D matrix where every row corresponds to a tile in the slide and the columns contain the features. The vectorized slide is the input to the GMA model which combines the tile representations into a slide-level representation which is then linearly projected to class scores.

To estimate generalization performance, we employed a Monte Carlo Cross-Validation (MCCV) strategy. For each MCCV split, 80\% of the samples were assigned to the training set and the remaining 20\% were assigned to the validation set. For each benchmark task, the 20 MCCV folds were randomly sampled and kept fixed for all experiments. Each MCCV split was run twice to assess stochastic fluctuations during training and the results were averaged across the two replicas. All models were trained using a single GPU for 50 epochs using the AdamW~\cite{loshchilov_decoupled_2017} optimizer. A cosine decay with warm up schedule was used for the learning rate with a peak learning rate of 0.0001. The exact parameters used for training can be found in the \href{https://github.com/fuchs-lab-public/OPAL/tree/main/SSL_benchmarks}{GitHub} repository. For each task and foundation model, the distribution of validation AUCs across the 20 MCCVs are used to assess the trained model performance.

\section{Downstream Tasks}\label{sec:tasks}
To assess the representation power of pathology foundation models, we collected a series of clinical datasets spanning clinically relevant tasks from multiple institutions and scanned with a variety of scanners. The tasks are described below and summarized in Table~\ref{tab:tasks_detection} and Table~\ref{tab:tasks_biomarker} for the detection and the biomarker tasks respectively.

\subsection{Disease Detection}
\begin{itemize}
    \item \textbf{MSHS Breast Cancer Detection Cohort.} Breast cancer blocks and normal breast blocks were obtained from the pathology LIS. A total of 1998 slides were sampled, with 999 positive and 999 negative. The positive slides were selected from blocks that received the routine biomarker panel for cancer cases (estrogen receptor ER, progesterone receptor PR, HER2, and Ki67), while negative slides were selected from breast cases that did not have an order for the routine panel. Additionally, negative cases were selected if they were not a mastectomy case, did not have a synoptic report associated with the case, and had no mention of cancer or carcinoma in the report.
    \item \textbf{MSHS Oral Cancers Detection Cohort.} Tumor (positive) and normal (negative) block information was extracted from structured synoptic reports obtained from the LIS. Synoptic reports for ``Lip and Oral Cavity'' were included. The positive samples included a variety of cancer diagnoses: squamous cell carcinoma, adenoid cystic carcinoma, mucoepidermoid carcinoma, and others.
    \item \textbf{MSHS Bladder Cancers Detection Cohort.} Tumor (positive) and normal (negative) block information was extracted from structured synoptic reports obtained from the LIS. Synoptic reports for ``Cystectomy, Anterior Exenteration'' and ``Transurethral Resection of Bladder Tumor'' were included. The positive samples included a variety of cancer diagnoses: urothelial carcinoma, small cell neuroendocrine carcinoma, adenocarcinoma, squamous cell carcinoma, and others.
    \item \textbf{MSHS Kidney Cancers Detection Cohort.} Tumor (positive) and normal (negative) block information was extracted from structured synoptic reports obtained from the LIS. Synoptic reports for ``Nephrectomy'' were included. The positive samples included a variety of cancer diagnoses: clear cell renal cell carcinoma, chromophobe renal cell carcinoma, papillary renal cell carcinoma, Xp11 translocation renal cell carcinoma, clear cell sarcoma, and others.
    \item \textbf{MSHS Thyroid Cancers Detection Cohort.} Tumor (positive) and normal (negative) block information was extracted from structured synoptic reports obtained from the LIS. Synoptic reports for ``Thyroid Gland'' were included. The positive samples included a variety of cancer diagnoses: papillary carcinoma, follicular carcinoma, Hurthle cell carcinoma, and others.
    \item \textbf{MSHS DCIS Detection Cohort.} Tumor (positive) and normal (negative) block information was extracted from structured synoptic reports obtained from the LIS. The synoptic report ``DCIS of the Breast'' was used for this cohort.
    \item \textbf{MSHS Prostate Cancer Detection Cohort.} Tumor (positive) and normal (negative) block information was extracted from structured synoptic reports obtained from the LIS. Synoptic reports for ``Radical Prostatectomy'' and ``Transurethral Prostatic Resection'' were included. The positive samples included acinar and ductal prostate adenocarcinomas.
    \item \textbf{MSHS Colo-rectal Cancers Detection Cohort.} Tumor (positive) and normal (negative) block information was extracted from structured synoptic reports obtained from the LIS. Synoptic reports for ``Resection'', ``Transanal Disk Excision of Rectal Neoplasms'', ``Excisional Biopsy (Polypectomy)'', and ``Neuroendocrine Tumor'' were included. The positive samples included a variety of cancer diagnoses: adenocarcinoma, signet-ring cell carcinoma, micropapillary carcinoma, and others.
    \item \textbf{MSHS IBD Detection Cohort.} Normal mucosa samples were obtained from patients undergoing screening and routine surveillance lower endoscopy from 2018 to 2022. Inflammatory bowel disease (IBD) cases, including first diagnoses and follow ups, were included. Active IBD samples were scored using the Mount Sinai histologic disease criteria and found to have Histologic Activity Score (HAI) $>= 1$.  A total of 1,441 slides were sampled, 717 with active inflammation and 724 with normal mucosa.
\end{itemize}

\begin{table}[h]
\caption{Summary of detection downstream tasks currently included.}\label{tab:tasks_detection}
\begin{tabular*}{\textwidth}{@{\extracolsep\fill}llrl}
\toprule
Origin & Disease & Slides (Positive) & Scanner \\
\midrule
MSHS & Breast Cancer & 1,998 (999) & Philips Ultrafast\\
MSHS & Oral Cancer & 279 (145) & Philips Ultrafast\\
MSHS & Bladder Cancer & 448 (272) & Philips Ultrafast\\
MSHS & Kidney Cancer & 1,000 (562) & Philips Ultrafast\\
MSHS & Thyroid Cancer & 710 (390) & Philips Ultrafast\\
MSHS & DCIS & 233 (135) & Philips Ultrafast\\
MSHS & Prostate Cancer & 1,000 (547) & Philips Ultrafast\\
MSHS & Colo-rectal Cancer & 413 (257) & Philips Ultrafast\\
MSHS & IBD & 1,448 (717) & Philips Ultrafast\\
\botrule
\end{tabular*}
\end{table}

\subsection{Computational Biomarkers}
\begin{itemize}
    \item \textbf{MSHS Breast Cancer ER Prediction Cohort.} Breast cancer cases with orders for Estrogen Receptor (ER) IHC were queried from the LIS. The IHC result was automatically extracted from the pathology report. A total of 2000 slides were sampled, 1000 positive, 1000 negative.
    \item \textbf{MSHS Breast Cancer PR Prediction Cohort.} Breast cancer cases with orders for Progesteron Receptor (PR) IHC were queried from the LIS. The IHC result was automatically extracted from the pathology report. A total of 1,986 slides were sampled, 953 positive, 1,033 negative.
    \item \textbf{MSHS Breast Cancer HER2 Prediction Cohort.}  Breast cancer cases with orders for HER2 IHC and FISH were queried from the LIS. IHC and FISH results were automatically extracted from the pathology report. A total of 2,018 slides were sampled, 760 positive, 1,258 negative.
    \item \textbf{MSHS Breast HRD Prediction Cohort.} Mount Sinai BioMe is a whole-exome sequencing cohort of 30k individuals, where carriers of pathogenic and protein-truncating variants affecting Homologous Repair Deficiency (HRD) genes (i.e., BRCA1, BRCA2, BRIP1, PALB2, RAD51, RAD51C, RAD51D, ATM, ATR, CHEK1, and CHEK2), where included as positives. A subset of the BioMe dataset of patients with available breast pathology slides were included. Slides containing solely normal breast tissue and slides with breast cancer were both included.
    \item \textbf{MSHS EGFR mutation detection LUAD.} Lung adenocarcinoma (LUAD) patients that underwent next generation sequencing (NGS) profiling for their cancer were identified. A total of 294 slides were obtained from MSHS’s clinical slide database, 103 positive and 191 negative. Mutations outside of the EGFR kinase domain (exons 18-24) are not considered oncogenic and are considered negative in this analysis. 
    \item \textbf{MSKCC EGFR Mutation Prediction in LUAD.} LUAD patients at Memorial Sloan Kettering Cancer Center with respective molecular analysis from the MSK-IMPACT assay~\cite{cheng_comprehensive_2017,zehir_mutational_2017} and corresponding digitized slides where identified. MSK-IMPACT is an NGS assay that can detect variants in up to 505 unique cancer genes, including EGFR. Mutations outside of the EGFR kinase domain (exons 18-24) are not considered oncogenic and are considered negative in this analysis. This is a sample of the dataset described in Campanella et al.~\cite{campanella_hampe-based_2022} where more information can be found. A total of 1,000 slides were sampled at random, 307 positive and 693 negative.
    \item \textbf{MSKCC TP53 Mutation Prediction in LUAD.} MSK-IMPACT derived TP53 mutational status. A total of 998 slides were sampled, 430 positive and 568 negative.
    \item \textbf{MSKCC KRAS Mutation Prediction in LUAD.} MSK-IMPACT derived KRAS mutational status. A total of 998 slides were sampled, 325 positive and 673 negative.
    \item \textbf{MSKCC STK11 Mutation Prediction in LUAD.} MSK-IMPACT derived STK11 mutational status. A total of 998 slides were sampled, 122 positive and 876 negative.
    \item \textbf{MSKCC ALK Mutation Prediction in LUAD.} MSK-IMPACT derived ALK mutational status. A total of 999 slides were sampled, 144 positive and 855 negative.
    \item \textbf{MSKCC ICI Therapy Response Prediction in NSCLC.} Non-small cell lung cancer (NSCLC) patients who received PD-L1 blockade-based immunotherapy between 2013 and 2019 at MSKCC were considered. Cytology specimens were excluded. Objective overall response was determined by RECIST and performed by a blinded thoracic radiologist. A total of 454 slides were obtained, 86 positive and 368 negative.
\end{itemize}

\begin{table}[h]
\caption{Summary of downstream tasks currently included for computational biomarker prediction.}\label{tab:tasks_biomarker}
\begin{tabular*}{\textwidth}{@{\extracolsep\fill}lllrl}
\toprule
Origin & Biomarker & Specimen & Slides (Positive) & Scanner \\
%\multicolumn{6}{@{}c@{}}{Disease Detection} \\
%\cmidrule{1-6}
\midrule
MSHS & IHC ER & Breast Cancer & 2,000 (1,000) & Philips Ultrafast\\
MSHS & IHC PR & Breast Cancer & 1,986 (953) & Philips Ultrafast\\
MSHS & IHC/FISH HER2 & Breast Cancer & 2,018 (760) & Philips Ultrafast\\
MSHS & BioMe HRD & Breast & 563 (188) & Philips Ultrafast\\
MSHS & NGS EGFR & LUAD & 294 (103) & Philips Ultrafast\\
MSKCC & NGS EGFR & LUAD & 1,000 (307) & Aperio AT2\\
MSKCC & NGS ALK & LUAD & 999 (144) & Aperio AT2\\
MSKCC & NGS STK11 & LUAD & 998 (122) & Aperio AT2\\
MSKCC & NGS KRAS & LUAD & 998 (325) & Aperio AT2\\
MSKCC & NGS TP53 & LUAD & 998 (430) & Aperio AT2\\
MSKCC & ICI Response & NSCLC & 454 (86) & Aperio AT2\\
\botrule
\end{tabular*}
\end{table}

\section{Foundation Models}\label{sec:models}
In this work we focus on benchmarking publicly available vision foundation models trained on large pathology corpora. These include: \href{https://github.com/Xiyue-Wang/TransPath}{CTransPath}~\cite{wang_transformer-based_2022}, \href{https://huggingface.co/MahmoodLab/UNI}{UNI}~\cite{chen_towards_2024}, \href{https://huggingface.co/paige-ai/Virchow}{Virchow}~\cite{vorontsov_virchow_2023}, and \href{https://huggingface.co/prov-gigapath/prov-gigapath}{Prov-GigaPath}~\cite{xu_whole-slide_2024}. We also include a truncated ResNet50 (tRes50) pretrained on ImageNet as a baseline due to its popularity in the computational pathology community. For Virchow, since it was trained on slides from MSKCC, we can't ensure that there is no overlap between their pretraining cohort and the clinical tasks based on MSKCC data. For Prov-GigaPath, the authors provide both a pretrained tile-level encoder and a pretrained slide-level aggregator, but since this work is focused on assessing the expressiveness of the feature representation of tile-level encoders, only the encoder portion of Prov-GigaPath is considered. For each foundation model, we followed the embedding instructions provided by the authors in each respective repository. 

For comparison, we further include two in-house trained foundation models: a ViT-small (21.7M parameters) and a ViT-base (85.8M parameters) trained with DINO~\cite{caron_emerging_2021}. These models were pretrained on a clinical dataset compiled at MSHS during normal hospital operation. The pretraining dataset consisted of 423,563 H\&E stained slides from 88,035 cases and 76,794 patients. These include slides from 42 organs across all pathology specialities. We ensured that no overlap exists between this pretraining dataset and the clinical benchmarking dataset. All slides were scanned on a Philips Ultrafast scanner at 40x magnification (0.25 MPP), de-identified and converted to tiff format. The total storage required for the raw tiff files was around 600TB. As a preprocessing step, tissue tiles were extracted from each slide at 0.5 MPP resolution, yielding approximately 3.2 billion tiles. The ViT-small (SP21M) was trained on 12 Nvidia A100 40GB GPUs with a batch size of 90 per GPU for 17 days and 16 hours. The ViT-base (SP85M) was trained on 8 Nvidia H100 80GB GPUs with a batch size of 100 per GPU for 26 days and 11 hours. Both models were trained on approximately 1.6 billion tiles. The models are publicly available on HuggingFace: \href{xxx}{SP21M}, \href{xxx}{SP85M}.

We can observe that older foundation models are trained with variants of contrastive learning. After the introduction of DINO, and later DINOv2, recent foundation models have used the latter as go-to pretraining algorithm. While evidence emerged that DINO tends to outperform contrastive learning and masked image modeling approaches for pathology pretraining~\cite{kang_benchmarking_2022,campanella_computational_2023}, there is no direct comparison of DINO and DINOv2. Since DINOv2 is considerably more computationally expensive than DINO, such a comparison would be desirable.

\section{Results}

\subsection{Disease Detection Tasks}

\begin{figure}[h]
\centering
\includegraphics[width=\textwidth]{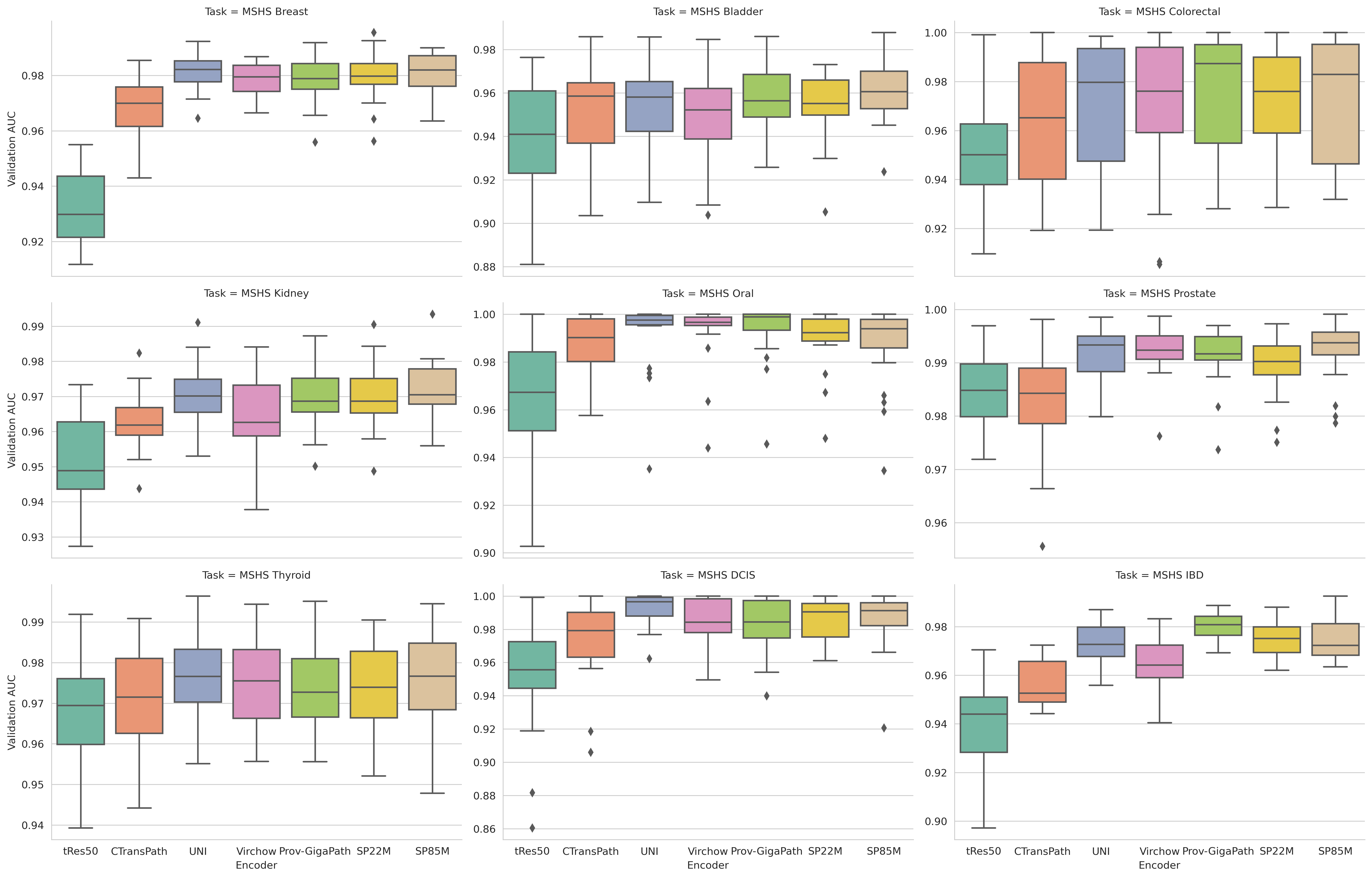}
\caption{Benchmarking Results: Detection Tasks.}\label{fig:detection}
\end{figure}

We present the results of the disease detection benchmarks in Figure~\ref{fig:detection}. The disease detection results show consistent performance across all tasks, independent of disease, with AUCs above 0.9 for all encoders tested. The ImageNet pretrained encoder is consistently under-performing the pathology trained encoders. Among the pathology trained encoders, CTransPath consistently shows inferior performance. In general, foundation model performance can be attributed to a combination of encoder architecture, pretraining dataset, and pretraining algorithm. CTransPath was trained on a small dataset and used a contrastive learning algorithm, which may explain the difference in performance. The other foundation models trained with DINO and DINOv2 show very similar performance despite differences in pretraining datasets and model architecture. The SP21M and SP85M models despite being much smaller and trained with DINO, achieve comparable performance to much larger models trained with DINOv2. Additionally, the composition of the pretraining dataset seems to play a lesser role for detection tasks. It is not well studied how to optimally design a pretraining dataset, including what is an acceptable size. In our analysis, UNI, with 100M pretraining tiles, is the most efficient of the models tested in terms of pretraining dataset size.
Overall, for detection tasks, all the DINO and DINOv2 trained models (SP21M, SP85M, UNI, Virchow, and Prov-GigaPath) achieve comparable performance and the choice of model may depend on other considerations, such as inference cost.

\subsection{Computational Biomarker Prediction Tasks}

\begin{figure}[h]
\centering
\includegraphics[width=\textwidth]{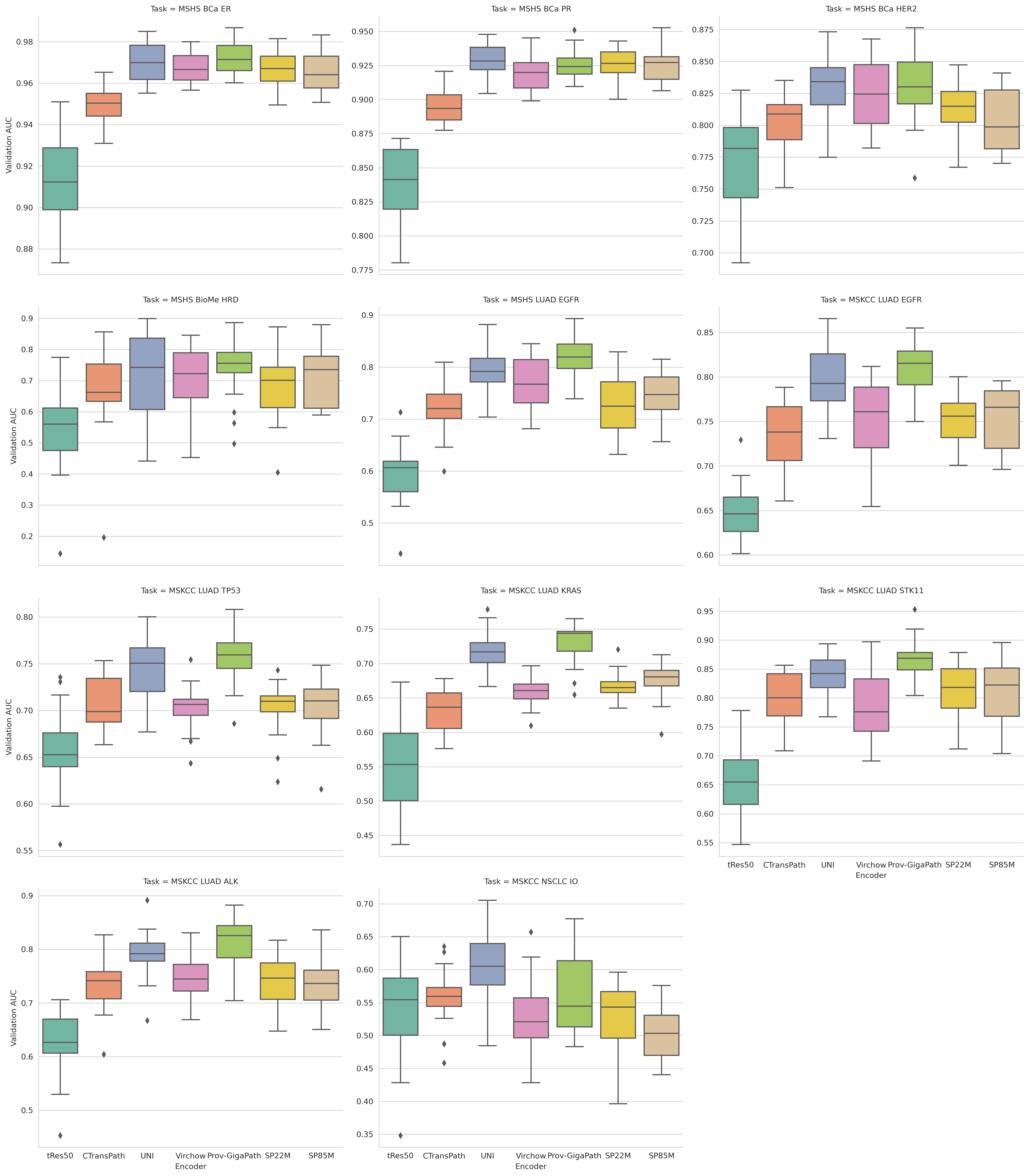}
\caption{Benchmarking Results: Biomarker Prediction Tasks.}\label{fig:biomarker}
\end{figure}

We present the results of the biomarker prediction benchmarks in Figure~\ref{fig:biomarker}.
Biomarker prediction tasks are more challenging than disease detection since it may be unknown whether a particular genomic alteration leads to a measurable morphological change visible in H\&E stained slides. For some biomarkers, prediction from H\&E may not be feasible. As expected, the biomarker prediction tasks show a higher degree of variability in performance than the detection tasks. The gap in performance of the ImageNet pretrained model is more evident here than in the detection tasks. Likewise, CTransPath tends to perform worse than the DINO and DINOv2 models as observed for the detection tasks.
For the biomarkers involving breast tissue, all DINO/DINOv2 models show similar performance. The main exception is the prediction of HER2 positivity, where UNI, Virchow, and Prov-GigaPath are superior than the DINO trained SP21M and SP85M.
For the NGS mutation alterations in lung, we observe that UNI and Prov-GigaPath achieve consistently better AUCs than Virchow, SP21M, and SP85M. These results may be explained by the fact that lung is over-represented in the pretraining datasets of UNI and Prov-GigaPath. For UNI, lung is the second most common tissue in their dataset, with around 10\% of the slides or about ten thousand WSIs. For Prov-GigaPath, lung is the most common tissue, comprising over 45\% of the slides, or about 77 thousand WSIs. In comparison, SP21M and SP85M were trained on about one thousand lung slides only.
This points to the hypothesis that while for detection tasks, dataset composition is less relevant, it may play a significant role for biomarker prediction. More specifically, for a subtype in question, e.g. lung, a higher percentage of that tissue in the training dataset might lead to a better representation of its variability in the final embedding and may explain the differences in the observed performance in these experiments.

Finally, for the task of predicting response to immune checkpoint inhibitors (ICIs) in non-small-cell lung cancer (NSCLC), all models obtained equally poor results with AUCs barely above chance. UNI, with and average AUC of 0.6 performed slightly better than others, but was still underwhelming. It is known that ICI response prediction from H\&E slides is a challenging task, yet there is evidence that descriptors of local cellular networks~\cite{xie_computational_2022}, that better model the tumor microenvironment (TME) can achieve AUCs of around 0.7, on-par with PD-L1 IHC, the current clinical gold standard. It is reasonable to hypothesize that SSL trained foundation models should be able to capture local cellular information and reach similar performance. One potential explanation is that the pretraining data may be skewed in terms of cancer presence, cancer subtypes, and cancer stage. Given that foundation models tend to be trained on large collections requiring less or no data curation, this level of detail is generally not available. Yet, this suggests that the composition of the pretraining dataset may be crucial, especially for challenging response prediction tasks.
Overall, for the biomarker prediction tasks, UNI and Prov-GigaPath were consistently as good or better than other models.

\subsection{Scaling Laws: Foundation Model Size}

We have shown that the composition of the pretraining dataset may be a crucial aspect for explaining downstream performance. Another key aspect is the representational capacity of the model which can be roughly comparable to the model's parameter count. Here we investigate the foundation model size to assess whether scaling laws observed in other domains such as natural language processing are occurring for pathology data. For this analysis we included only vision transformers trained with DINO or DINOv2. Model sizes range from 22 million (SP22M) to 1.1 billion (Prov-GigaPath) parameters.
%Dataset sizes range from 100 million (UNI) to 1.6 billion (SP22M and SP85M) tiles and 100 thousand (UNI) to 1.5 million (Virchow) slides.
The complete information is curated in Table~\ref{tab:works}.

\begin{figure}[h]
\centering
\includegraphics[width=\textwidth]{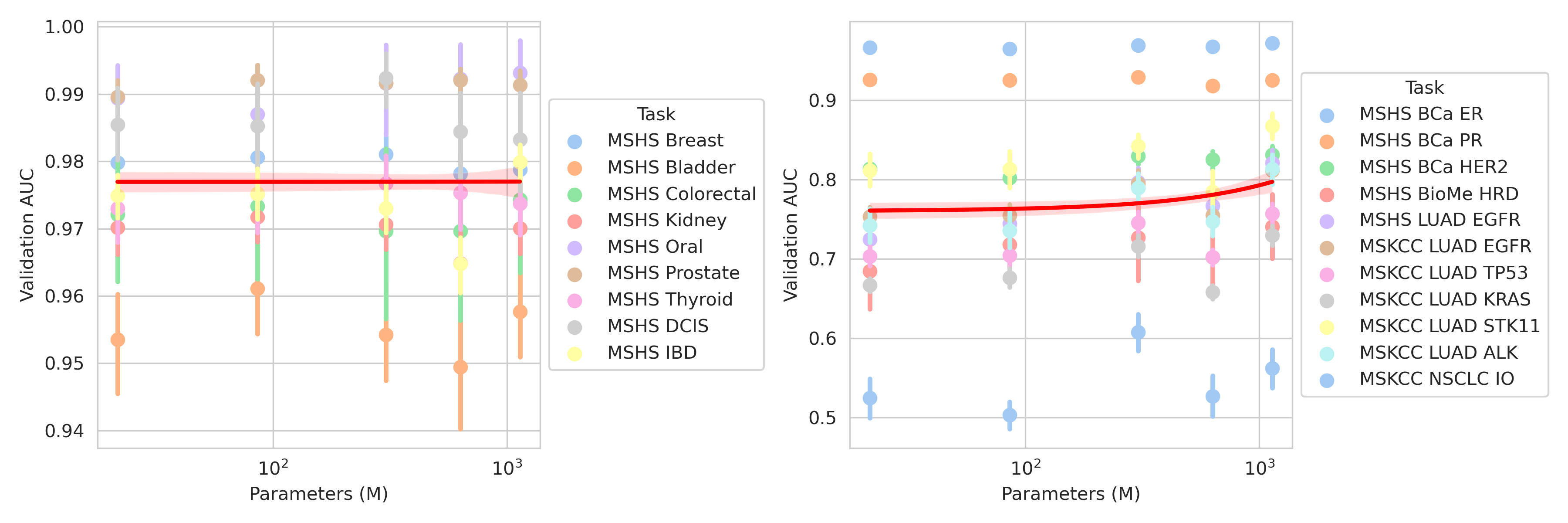}
\caption{Scaling Laws: downstream performance vs foundation model size.}\label{fig:size}
\end{figure}

Figure~\ref{fig:size} shows how the downstream performance of detection and biomarker prediction tasks correlate with encoder model sizes. For detection tasks, our results suggest that there is no evidence of downstream performance scaling with model size. As we showed previously, on average a 22 million parameter model is comparable to a 1.1 billion parameter model for these tasks.
In contrast, for biomarker prediction, an overall tendency of higher performance with larger models is observed. Yet, we caution that these conclusion depends on the tasks currently included in the analysis and it may be better to assess each task independently. For several breast biomarkers, there is no benefit from larger models, whereas for the NGS lung tasks there seems to be a benefit. Yet, as we described before, this may be due to the pretraining dataset composition and not the larger model capacity.

\subsection{Scaling Laws: Computational Resources}
A key aspect of training foundation models is that of computational resources. While in the general computer vision community there are a number of private efforts to train very large scale (in terms of both model and data size) vision foundation models, in the medical domain, and in pathology particularly, resources are still far more scarce. Academic and private research groups alike have published models trained for the most part with much more modest computational resources. Here we analyze the overall computational resources used to pretrain public foundation models and how they correlate with downstream performance.

Computational resources needed to train a model depend on the model architecture, dataset size, and algorithm used for pretraining. Model architecture influences the GPU memory needed to process a batch of samples: smaller models can fit much larger batches than larger models. Dataset size is directly related to the number of optimization steps performed during training, especially in large scale SSL experiments where it is common that each data point is used only once.
Finally, the algorithm defines the data augmentations and computations performed and is therefore a key aspect to consider.
%Here we focused on the DINO and DINOv2 trained models. 
Compared to other SSL algorithms, DINO and DINOv2 are more computationally expensive, with DINOv2 being the more expensive one of the two.
For example, SP22M with 22 million parameters was trained with DINO using full precision on 40GB GPUs with a batch size per GPU of 90 tiles. In comparison, Prov-GigaPath with 1.1 billion parameters was trained with DINOv2 using half precision on 80GB GPUs with a batch size of 12 tiles per GPU.

To measure overall computational resources we use GPU-hours but normalize it to a hypothetical 80GB GPU card. We assume that for models trained on a 40GB card, the computation time would be halved by using an 80GB card. GPU usage and training times were obtained from each respective paper, model cards in public repositories, or by personal correspondence and are summarized in Table~\ref{tab:GPUs}.

\begin{table}[h]
\caption{Computational resources for the pretraining of public pathology foundation models considered in this analysis.}\label{tab:GPUs}
\begin{tabular*}{\textwidth}{@{\extracolsep\fill}lrlrrr}
\toprule
Model & \multicolumn{2}{c}{GPU} & Batch Size & Training Time & Feature \\
 & N & Type & Per GPU & (hours) & Length\\
\midrule
%CTransPath~\cite{wang_transformer-based_2022} & 48 & V100 & 96 & 250 & \\
%Phykon~\cite{filiot_scaling_2023} & 32 & V100 & 45 & 1,216 & \\
UNI~\cite{chen_towards_2024} & 32 & A100-80GB & 96 & 32 & 1,024\\
Virchow~\cite{vorontsov_virchow_2023} & NA & A100-40GB & 16 & NA & 2,560\\
SP22M & 12 & A100-40GB & 90 & 424 & 384\\
SP85M & 8 & H100-80GB & 100 & 635 & 768\\
%Rudolf-V~\cite{dippel_rudolfv_2024} & 16 & A100-40GB & 960 & N/A & \\
Prov-GigaPath~\cite{xu_whole-slide_2024} & 32 & A100-80GB & 12 & NA & 1,536\\
\botrule
\end{tabular*}
\end{table}

\begin{figure}[h]
\centering
\includegraphics[width=\textwidth]{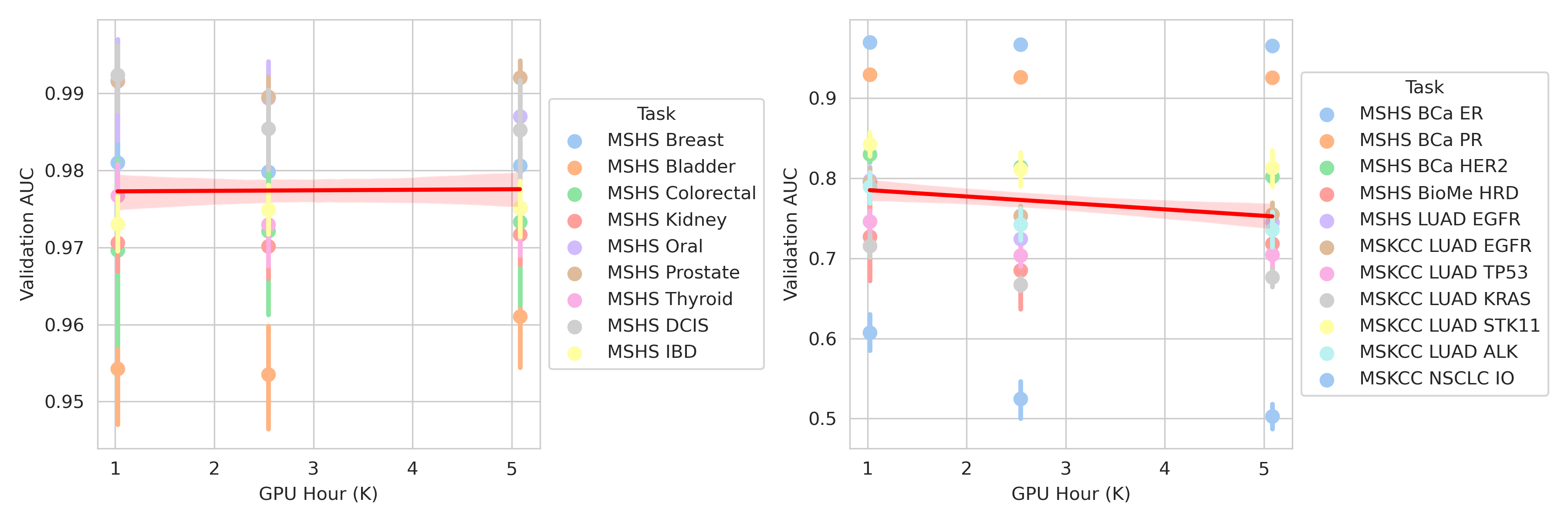}
\caption{Scaling Laws: downstream performance vs computational resources used for pretraining the foundation models.}\label{fig:resources}
\end{figure}

Figure~\ref{fig:resources} shows how the downstream performance of detection and biomarker prediction tasks correlate with computational resources used for training. We were able to quantify resources only for our in-house models and UNI. 
For detection tasks, our results show no evidence of improved performance associated with higher computational costs. The same conclusion can be made for biomarker prediction tasks, where UNI, which, despite being a larger model, overall used modest computational resources performs better on average than our SP22M and SP85M models which used significantly more computational resources.

\section{Discussion}

Self-supervised learning and foundation models have the potential to revolutionize medical research. Early work in computational pathology is showing a clear benefit over traditional supervised approaches in terms of performance and generalizability. Notably, recent models trained by both academic and private institutions are being released in public repositories, empowering researchers with powerful tools to develop the next generation of predictive models. While there is still much work to be done towards democratizing these technologies, and making them available to the research community, the emergence of foundation models may play a significant role.

As more and more foundation models are trained, an independent benchmark of clinically relevant tasks becomes essential for both researchers training foundation models and for researchers looking to apply these pretrained foundation models on downstream tasks. Training new foundation models is expensive and it is important to learn from previous efforts. A benchmark can provide insights for improving pretraining and yield better models in the future. For downstream clinical applications, a benchmark can guide the decision to use one model over another, considering a variety of factors, from performance on various tasks to computational resource constraints. In this work, we presented a benchmark of publicly available pathology foundation models focusing on twenty clinically relevant slide level tasks across a variety of tissues and disease indications. Importantly, all the data was generated during clinical operations without further curation, representing the variability both biological and technical that can be observed under real world conditions.

We made a deliberate decision to not release the test data used for these benchmarks. Efforts to scrape all publicly available data for pretraining foundation models may lead to data contamination and negatively impact the relevance of such benchmark results. Instead, we will make an effort to regularly update the benchmark results with more models as they become publicly available and more downstream tasks. In the future, we plan to provide an API for users to benchmark their external models on our tasks.

In summary, the ImageNet pretrained encoder, and CTransPath to a lesser degree, consistently underperform compared to newer models. In disease detection and biomarker prediction tasks, all DINO and DINOv2 trained models performed comparably, with UNI and Prov-GigaPath models achieving better results in certain biomarker tasks on tissue that was overrepresented in their training datasets. The size of the model did not significantly impact disease detection performance, but larger models tended to perform better in biomarker prediction. However, these benefits varied depending on the task and could be influenced by the composition of the pretraining dataset.

From our analyses, we can make the following observations:
There is not yet strong evidence that scaling laws observed in pretraining SSL models for natural language and images are applicable for tile encoders in pathology.
Performance does not appear to scale with model size and dataset size as in other domains given current training algorithms.
Smaller models perform on par with much larger models on most tasks and are only marginally worse in others.
Similarly, the dataset size and overall computational expense does not appear to lead to significantly better models.
While there is unanimous agreement that DINO and DINOv2 perform better than other popular SSL algorithms for pathology pretraining, there is not much evidence to support a choice between the two.
There is some evidence that dataset composition may be a crucial aspect in the downstream performance, and more efforts in the curation of the pretraining data is likely to be beneficial. While general purpose foundation models may be desirable, tissue specific foundation models may be a viable alternative.
Yet, with current algorithms we can expect incremental performance gains, and likely not great leaps forward as exemplified by the ICI response task. It may be that we are reaching a limit to how much relevant histological information can be learned via SSL alone with current strategies.

%Things that are still unclear.
There are several aspects of pretraining a pathology foundation models that we could not address at this time due to lack of evidence.
%Magnification
The majority of foundation models have been trained at 20x magnification as it allows to use the largest possible cohort of data. One question is whether a higher resolution may be beneficial especially for tasks where cellular features may be important. Some works have started to appear where 20x and 40x magnification are used jointly. Whether mixing magnifications or training magnification specific models is of an advantage is largely unanswered.
%Stains
Similarly, a majority of efforts have focused on using H\&E stained slides and ignoring IHC ones. H\&E slides are the basis of diagnostic work and are the fastest and cheapest to produce. Meanwhile, IHC slides provide supporting information but are slower and more costly to generate and are used more sparingly. Since most computational pathology studies focus on predicting various endpoints directly from H\&E, it is reasonable that they have been the focus on pretraining foundation models. As a result, this seems to make foundation models less useful for IHC-based computational pathology models. Furthermore, it is possible that the inclusion of IHC would be beneficial for H\&E based tasks as well. Future work will be needed to address these questions.
%Multicenter v single center
Finally, gathering large collections of pathology slides for pretraining is a daunting task within the constraints of single institutions. While, collecting multi-institutional pretraining data might improve the robustness and generalizability of foundation models, there are several important obstacles in the way, and it has yet to be proven beneficial or necessary.

%Need a closing paragraph.

%\begin{figure}[h]
%\centering
%\includegraphics[width=0.9\textwidth]{fig.eps}
%\caption{This is a widefig. This is an example of long caption this is an example of long caption  this is an example of long caption this is an example of long caption}\label{fig1}
%\end{figure}

\backmatter

%\bmhead{Supplementary information}

%This article contains Supplementary Information.

\bmhead{Acknowledgements}

This work is supported in part through the use of the AI-Ready Mount Sinai (AIR.MS) research platform and the expertise provided by the team at the Hasso Plattner Institute for Digital Health at Mount Sinai (HPI.MS).

This work was supported in part through the computational and data resources and staff expertise provided by Scientific Computing and Data at the Icahn School of Medicine at Mount Sinai and supported by the Clinical and Translational Science Awards (CTSA) grant UL1TR004419 from the National Center for Advancing Translational Sciences.

\section*{Declarations}

\subsection*{Code Availability}
Code used for pretraining SP22M and SP85M was taken from the official DINO repository. 
Code to run benchmarks will be made available upon acceptance.

\subsection*{Data Availability}
Benchmark data is deliberately not made available due to the potential risk for data leakage in the training of future foundation models. The benchmark will be continuously updated with available public foundation models.

\bibliography{references}% common bib file
%% if required, the content of .bbl file can be included here once bbl is generated
%%\input sn-article.bbl

%%%%%%%%%
%Cover letter
%Authors should provide a cover letter that includes the affiliation and contact information for the corresponding author. Authors should briefly discuss the work's importance and explain why the work is considered appropriate for the diverse readership of Nature Communications. Authors are asked to provide the names and contact information for qualified scientific reviewers and they may request the exclusion of certain referees. Finally, authors should indicate whether they have had any prior discussions with a Nature Communications editor about the work described in the manuscript.
%%%%%%%%%%

\end{document}